# Highly Conductive PDMS/Carbon Nanotube/Graphene Nanocomposites and Effect of Sulfuric Acid Treatment on Conductivity

Abuzer Alp Yetisgin[1,5,*], Hazal Sakar[2,5,6], Hakan Bermek[3], Levent Trabzon [2,4,5,6]

[1] Faculty of Engineering and Natural Sciences, Materials Science and Nano-Engineering Program, Sabanci University, 34956, Istanbul, Turkey; yalp@sabanciuniv.edu (A.A.Y.)
[2] Department of Nanoscience and Nanoengineering, Istanbul Technical University, 34469, Istanbul, Turkey
[3] Department of Molecular Biology and Genetics, Istanbul Technical University, 34469, Istanbul, Turkey; bermek@itu.edu.tr (H.B.)
[4] Faculty of Mechanical Engineering, Istanbul Technical University, 34469, Istanbul, Turkey
[5] Nanotechnology Research and Application Center - ITUnano, Istanbul Technical University, Istanbul, Turkey; sakare19@itu.edu.tr (H.S.); levent.trabzon@itu.edu.tr (L.T.)
[6] MEMS Research Center, Istanbul Technical University, 34469, Istanbul, Turkey
[*] Correspondence to: Abuzer Alp Yetisgin (yalp@sabanciuniv.edu) (Currently at Sabanci University)

**Abstract**

Poly(dimethylsiloxane) (PDMS) is an insulator and it is commonly used in fabrication of micro-structures but there is a need of electrical sonductive pdms. Therefore, this research is mainly focused on producing PDMS nanocomposites with high electrical conductivity. Thus, we produced highly conductive PDMS nanocomposites which were filled with multi-walled carbon nanotubes (MWCNTs) and graphene nanoplatelets (GNPs). Due to the synergistic effect between CNTs and GNPs inside a polymeric matrix, we expected to obtain more conductive nanocomposites than PDMS nanocomposite filled with only CNTs. Additionally, we investigated the effect of sulfuric acid treatment to surface composition and electrical conductivity of prepared PDMS/MWCNT/GNP nanocomposites. Results indicated that electrical conductivity of sulfuric acid-treated samples was significantly higher than untreated samples. The level of conductivity in the range of 270.7 – 1074.9 S/m was obtained, and the higher ones are the samples treated with acid solution.

# 1. Introduction

Poly(dimethylsiloxane) (PDMS) is an elastomer which is used as adhesives or sealants in the industry like coating the electronic devices, and more recently, they have been used in applications of photovoltaics and aerospace industry.[1] PDMS has some interesting aspects including flexibility, optical transparency, biocompatibility, chemical stability, high resistivity, and hydrophobicity.[2,3] Furthermore, the advantageous properties of PDMS are its low cost, ease of fabrication of micro- or nano-structures by cast molding technique, and it could be able to seal other flat surfaces without using any adhesives.[3] Because of these properties, PDMS has been one of the leading polymer which is mostly used in microfluidics, microdevices, and micro-electrical mechanical systems (MEMS).[4,5] In spite of these qualities of PDMS, the insulator nature of PDMS limits some of its applications in these fields, because often PDMS-based devices require some degree of conductivity.[2] The flexibility of the PDMS makes it difficult to deposit or coat metal onto the surface of PDMS.[6] That is the reason why most of the studies are focused on increasing the conductivity of PDMS by adding conductive nanofiller inside the polymer matrix. High electrical conductivity and the flexibility of a nanocomposite will be able to generate new applications or improve some current applications in utilizing flexible conductive materials including strain and pressure sensors [7,8], conductive inks [9], gas sensors [10], biosensors [11–13], electrocardiograms [14,15], wearable devices [16], and so on.[17]

Different kinds of nanofiller materials are used for increasing conductivity of PDMS including carbon black (CB), multi-walled carbon nanotubes (MWCNTs), single-walled carbon nanotubes (SWCNTs), graphene nanoplatelets (GNPs), and metallic nano- or microparticles.[5] Among them, carbon nanotubes (CNTs) have high electrical and thermal conductivity, good thermal stability and high aspect ratio.[1,18] Polymer nanocomposites, which are utilized CNTs as nanofillers, have shown increased electrical conductivity by several orders of magnitude at a very low percolation threshold i.e. the minimum amount of nanofiller that is required for DC current to flow through polymer.[19–21] Nonetheless, GNPs are also highly conductive nanofillers but the percolation threshold of GNPs is higher than CNTs, the reason for that is, GNPs have a non-distinctive aspect ratio but CNTs have distinctive aspect ratio.[1,22] Despite all these, high amount of CNTs, or GNPs are needed to be added inside PDMS to make nanocomposites conductive enough e.g. Liu and Choi prepared PDMS/MWCNT nanocomposites with different MWCNT concentrations, and the conductivity of PDMS/MWCNT nanocomposites are found to increase from 0.003 S/m with 3 wt% of MWCNT to 6.3 S/m with 15 wt% of MWCNT.[2] Lee et al. combined PDMS and MWCNTs by using the solution mixing method and,

PDMS/MWCNT nanocomposites achieved electrical conductivity of 62,9 S/m after loaded with 25 wt% of MWCNTs.[8] Furthermore, MWCNTs do not just improve the electrical conductivity, they also increase the viscosity of nanocomposites which makes it harder to design fine structures by these PDMS/MWCNT nanocomposites. Therefore, optimum nanofiller concentration must be determined for the level of the finest of the structures that will be prepared by the nanocomposites. Recently, combinations of more than one type of nanofillers are added to polymer matrixes which are called polymer/hybrid nanocomposites.[23] Especially, CNT and GNP hybrid nanocomposites are the subject of attention. When CNTs and GNPs are incorporated into a polymer matrix, they tend to self-assemble due to π-π interactions which are inhibiting aggregation of nanofillers.[24] Kong et al. were prepared PDMS hybrid nanocomposite by using exfoliated graphene nanoplatelets (xGNP) and hydroxyl functionalized MWCNT (MWCNT-OH), which has higher electrical conductivity than nanocomposites prepared by MWCNT-OH or xGNP individually.[23] Similarly, Oh et al. were prepared a hybrid nanocomposite system that was made by mixing thermally reduced graphenes (TRGs) and MWCNTs. In both of these studies demonstrated that PDMS hybrid nanocomposites surpass the electrical conductivity of PDMS/MWCNT or PDMS/graphene nanocomposites.[25]

A number of different methods have been utilized for the preparation of polymer/CNT nanocomposites which include solution mixing, ball milling, melt mixing, solution casting, and so on. The common trait of all of these methods is, well dispersing of the CNTs inside the polymer matrix.[20,26] In order to fabricate well-improved nanocomposites with CNTs, good alignment of nanofillers, and aspect ratio need to be considered.[27] The most common nanocomposite preparation method is the solution mixing, which is done by dispersing CNTs (or other kinds of nanofiller materials) and polymer in some solvent, then it would be seen, the solvent is evaporated completely.[28] Choosing an appropriate solvent is extremely important in the solution mixing method. Because the method is relatively easy and has only a few modifiable parameters like agitation or stirring time of solution, or evaporation time of solvent, which is actually linked to the choice of solvent. Often chloroform, toluene, hexane, tetrahydrofuran (THF) and dimethylformamide (DMF) were used as solvents.[18,29–31] Highly volatile solvents e.g. chloroform could not be practical when the mixing time of nanocomposite solution is long.[32] On the other hand, apolar solvents like toluene and hexane also could not be usable, because these solvents could swell the PDMS.[29,32] Kim et al. demonstrated that the isopropyl alcohol (IPA) as a solvent has good dispersion quality of CNTs and PDMS, also partially soluble in IPA.[32] Ramalingame et al. showed dispersion qualities and stabilities of the different solvent including IPA, THF, toluene, and chloroform. As for the result, IPA and THF have nearly the same dispersion quality and stability.[31] Alcohol solvents have hydrophobic region and hydrophilic region which enables them to disperse CNTs inside the nanocomposite matrix.[32]

In this study, we demonstrated the synergistic effect of MWCNTs and GNPs inside the PDMS matrix for measuring the increase in electrical conductivity between PDMS/MWCNT and PDMS/MWCNT/GNP nanocomposites. We also investigated the effect of sulfuric acid treatment on the electrical conductivity of the nanocomposites. To do this, we etched the surface of the nanocomposites with sulfuric acid for various time periods and measured the electrical conductivity. For controlling the etching of nanocomposites, we used sulfuric acid with concentration of 90% $\pm$ 1,8. Furthermore, the solution mixing method is also applied for the production of PDMS/MWCNT and PDMS/MWCNT/GNP nanocomposites. IPA is preferred for the solvent for the fabrication of the nanocomposites, because of the mixing time of the nanocomposite solutions. This study made with the aim of producing PDMS/MWCNT/GNP nanocomposite with high electrical conductivity which will be able to be utilized as electrode material for further studies.

## 2. Experimental

### 2.1. Chemicals and Materials

Polydimethylsiloxane (PDMS, slygard-184 silicone elastomer) was purchased from Dow-Corning. Multi-walled carbon nanotubes (MWCNTs, >90%, were grown by chemical vapor deposition with 10-15 nm diameter, and the length of 3 µm), and graphene nanoplatelets (GNPs, >99.5%, with 6 nm thickness, the surface area of 150 $m^2/g$, and the diameter of 24 µm) were purchased from Nanografi Co. Ltd. (Ankara, Turkey). Isopropyl alcohol (IPA), sulfuric acid (95-97%) and other chemicals are purchased from Sigma Aldrich.

### 2.2. PDMS/CNT/GNP Nanocomposite Preparation

MWCNT/GNP nanofillers added to beherglass with different ratios including 1:6, 2:5, 1:1, 5:2, 6:1 and 7:0 (7 wt% to PDMS), and these nanocomposites were named as PDMS16, PDMS25, PDMS11, PDMS52, PDMS61 and PDMS70, respectively. These nanofiller concentrations mixed with the appropriate amount of IPA, and homogeneously dispersed in the ultrasonic bath for 30 min. Then, PDMS was weighed and put it into MWCNT/GNP/IPA dispersions. After PDMS addition, nanocomposite solutions were mixed under mechanical stirrer at 1000 rpm for

30 min. For removing IPA from nanocomposite mixtures, homogeneously dispersed PDMS/MWCNT/GNP/IPA mixtures were put inside the vacuum oven and baked at 100 ºC for overnight. As prepared PDMS/MWCNT/GNP nanocomposite mixtures were cured by adding PDMS curing agent to nanocomposite mixtures (PDMS nanocomposite to curing agent weight ratio was 10:1). PDMS/MWCNT/GNP nanocomposites were cured at 120 ºC for 30 min.

For the measurements of the effect of the sulfuric acid etching to the electrical conductivity of PDMS nanocomposites, one side of cured PDMS61 nanocomposites was treated by 90% $\pm$ 1,8 sulfuric acid solution with different intervals of 5, 10, 15 and 30 seconds, and were called PDMS5s, PDMS10s, PDMS15s, PDMS30s, respectively. Cured empty PDMS sample also treated with sulfuric acid for 30 s, which was called PDMS-H2SO4.

## 2.3. Electrical Conductivity Measurements of PDMS Nanocomposites

Cured PDMS16, PDMS25, PDMS11, PDMS52, PDMS61, PDMS70, PDMS5s, PDMS10s, PDMS15s, PDMS30s samples are cut into 2 x 1 cm$^2$ rectangular pieces. The thickness of the samples was measured by a digital micrometer. Thickness measurements are made at three different points and the mean thickness of samples was calculated for conductivity measurements.

Electrical conductivity measurements of the samples were calculated from sheet resistance, Rs, (ohm.sq) values obtained by JG M3 mini four-point probe ST2253 (Suzhou, China). Firstly, real sheet resistance values calculated by utilizing the coefficients of thickness and shape of the samples which are given in the four-point probe's user manual. After that, for calculation of resistivity, ρ, values (Ohm.m) were done by multiplying Rs values with thickness (m) of individual samples, and bulk conductivity, σ, values (S/m) were calculated by eq. 1 and eq. 2, respectively. Equations used for bulk conductivity values was shown below.

$$\rho = Rs * thickness \tag{1}$$

$$\sigma = \frac{1}{\rho} \tag{2}$$

## 2.4. Characterization of PDMS Nanocomposites

For Scanning Electron Microscopy (SEM) analysis, PDMS samples (PDMS70, PDMS61, PDMS5s, PDMS10s, PDMS15s, and PDMS30s) are cleaned with ethanol and rinsed with deionized water. Samples are coated with the Au-Pd layer. Only the surface of the samples was analyzed by Tescan Vega-3 SEM instrument.

Samples of PDMS70, PDMS61, PDMS5s, PDMS10s, PDMS15s, and PDMS30s are investigated by the attenuated total reflection-Fourier transform infrared spectroscopy (ATR-FTIR) spectroscopy (ATR-FTIR). The spectra are obtained from Perkin-Elmer Spectrum 100 ATR-FTIR spectrophotometer.

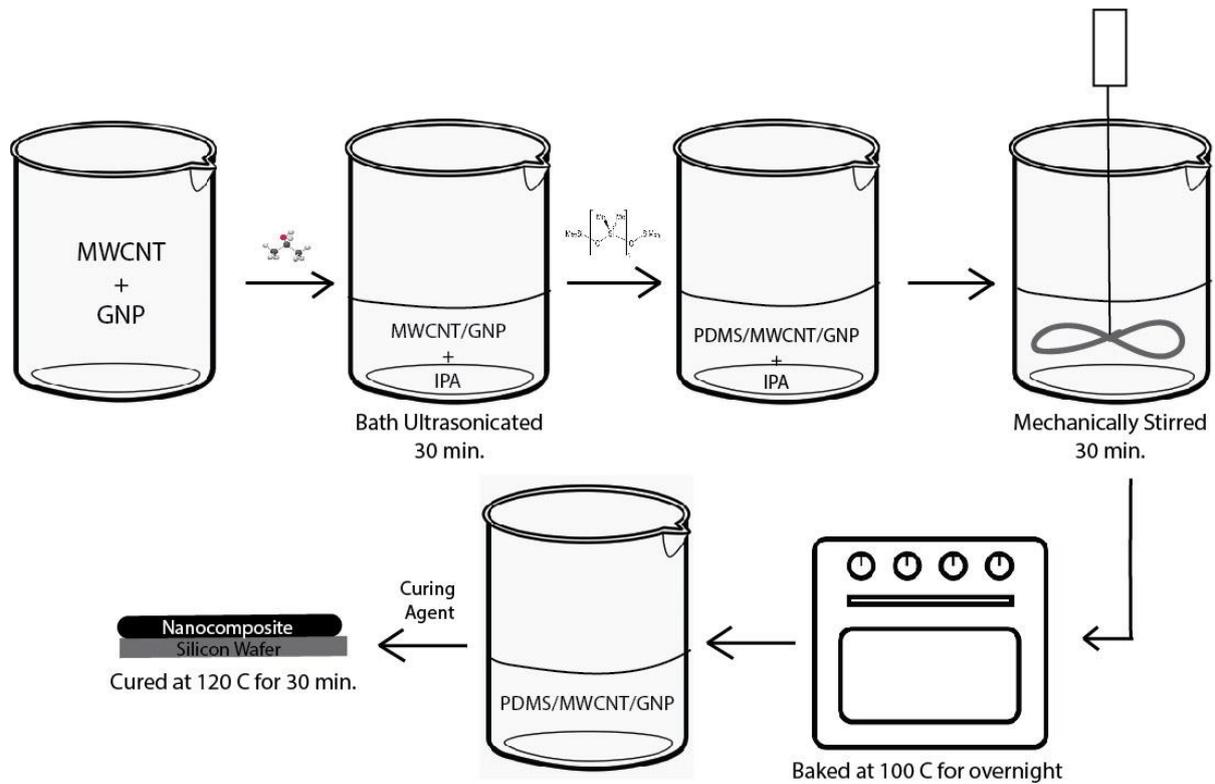

**Figure 1:** Schematic representation of PDMS/MWCNT/GNP nanocomposite preparation

## 3. Results and Discussion

The preparation of the PDMS/MWCNT/GNP nanocomposites was done by an easy 5 step solution mixing method (see Fig. 1). One of the key steps in our method is the completely dispersing MWCNTs and GNPs inside IPA, therefore this step hampers the bundling of MWCNTs, or restacking of GNPs. IPA is used as solvent for previous studies that worked on the preparation of conductive nanocomposites.[32–34] We used IPA as solvent because it is relatively less volatile than polar aprotic solvents such as chloroform, or dimethylformamide (DMF), which is important for cost efficiency. Moreover, IPA could partially dissolve both MWCNTs and PDMS.[32] In the research, we first tried to increase electrical conductivity of the PDMS by adding only MWCNTs inside, however, the electrical conductivity of prepared PDMS/MWCNT nanocomposites did not enough for low amount of MWCNT, and when we increased the amount of MWCNT, the increased viscosity of the nanocomposite made it difficult to mold it. Furthermore, the cost of the production of the nanocomposite increased proportionally to MWCNT amount. In order to overcome these shortcomings, relatively cheap GNPs are added. This kind of nanocomposites with hybrid nanofillers already proved its efficiency for increasing electrical properties of the polymer matrix they were added when compared to nanocomposites with only one kind of nanofiller.[14,25,35]

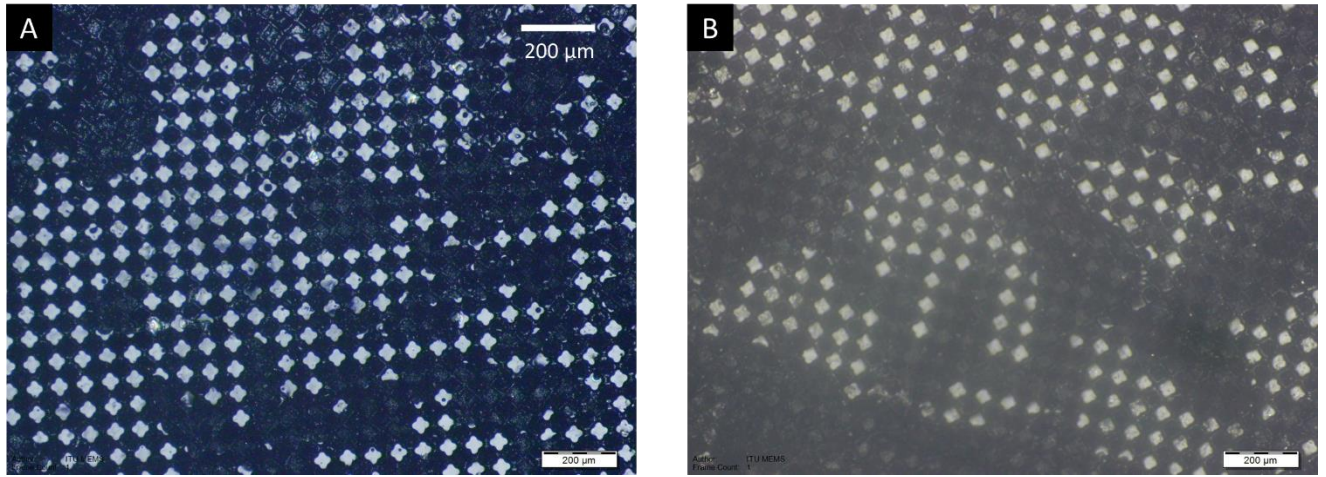

**Figure 2.** Optical microscopy images of micro-structured samples. A) PDMS61, B) PDMS5s

As mentioned before, viscosity of nanocomposites was increased with increasing nanofiller concentration. In Fig. 2, optical microscopy (Olympus, BX61) images of nanocomposites with micro-structured motifs were shown. Micro-structured motif consisted of micro-columns with 80 µm in height and 40 µm in width. Due to the high viscosity of PDMS61 nanocomposite, some areas of the nanocomposite did not take shape of micro-structured motif as shown in Fig. 2A. Moreover, after sulfuric acid treatment, microstructured motif deformed further, due to etching of sulfuric acid (see Fig. 2B). Liu and Choi explained that the higher conductivity was obtained by increasing nanofiller concentration, but nanocomposite turned to be too viscous for spin-coating.[2]

Over the course of our research, we observed that the sulfuric acid-treated surface of PDMS/MWCNT/GNP nanocomposite had higher conductivity as compared to untreated nanocomposite. Moreover, we even observed an increase in electrical conductivity of pristine PDMS samples after treated with sulfuric acid which showed in Fig 5. A number of studies have been done to see sulfuric acid treatment on the PDMS surface to record changes in its contact angle, or even selectively etching PDMS for constituting micropatterns on top of the PDMS surface.[36,37] However, we did not find any similar results in the literature on the effect of sulfuric acid treatment to the electrical conductivity of the nanocomposite surface.

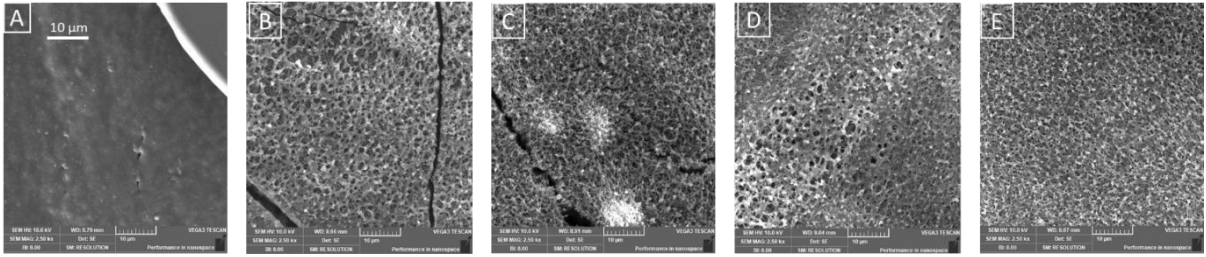

**Figure 3.** SEM images of the surfaces of the untreated and $H_2SO_4$ treated PDMS/CNT/GNP nanocomposite. A) Untreated PDMS/CNT/GNP, B) PDMS5s, C) PDMS10s, D) PDMS15s, E) PDMS30s

In the SEM images of the sulfuric acid-treated samples, which are found in Fig. 3, when the treatment time increases, the texture of the surface disrupts even further. Additionally, with the increasing treatment time, sulfuric acid penetrates into deeper parts of the sample, and degrades PDMS more. Besides, white precipitates were observed after the sulfuric acid treatment. Previous study explained these white precipitates, which were low molecular weight oligomers of PDMS.[37] SEM image, as shown in Fig. 4. demonstrated the distribution of MWCNTs and GNPs inside the nanocomposite matrix. MWCNTs and GNPs constructed a pathway that conducts electricity along the nanocomposite.

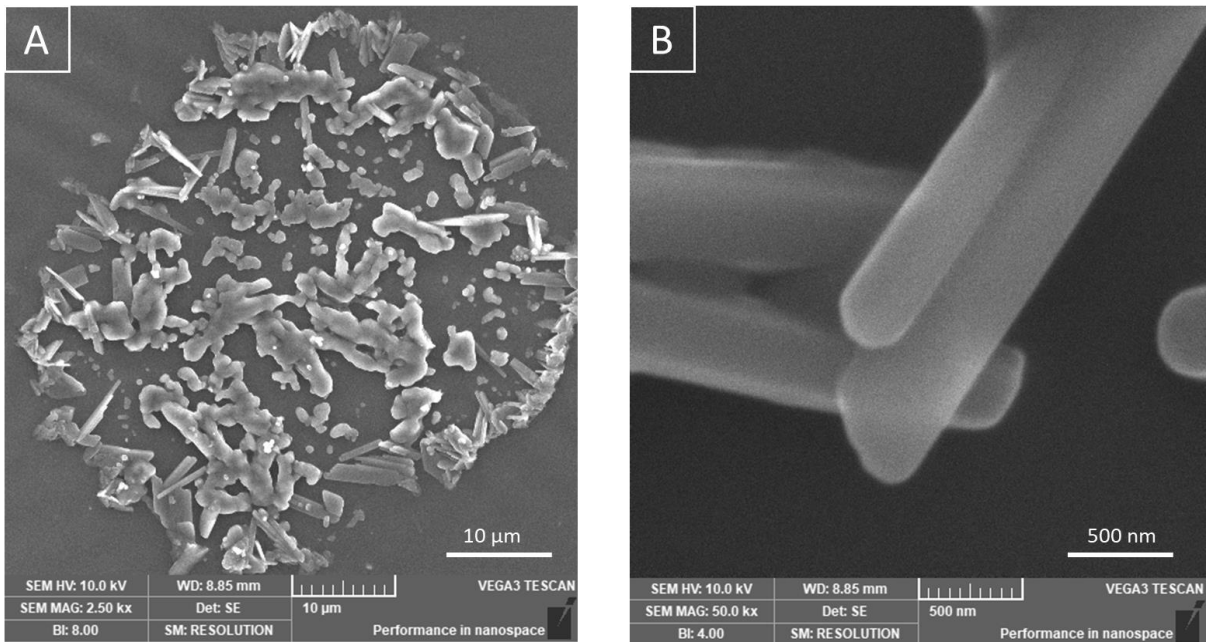

**Figure 4.** Sem images of the PDMS/CNT/GNP nanocomposite. A) Distribution of the CNTs and GNPs in the nanocomposite. B) CNTs and GNPs formed a junction inside nanocomposite.

For evaluation of structural changes occurred to samples after sulfuric acid treatment, two FTIR analysis were utilized for the two types of samples. The first one was made for investigating the difference of the composition between pristine untreated PDMS and the PDMS that treated with sulfuric acid for 30 s. The second one was made for analyzing the difference between untreated and treated PDMS/MWCNT/GNP nanocomposites nanocomposites as shown in Fig. 5. In the FTIR data, there is broad peak between the wavelength of 3000 to 3600 cm$^{-1}$ that appeared after sulfuric acid treatment, this peak corresponded to hydroxyl group that formed with the sulfuric acid treatment. Moreover, the peaks at 910 cm$^{-1}$, 1176 cm$^{-1}$, and 1560-1830 cm$^{-1}$ advocate the existence of oxygen single or double bonds. In the Fig. 5A, bands at 900-930 cm$^{-1}$ suggest that the transmission value decreased with increasing MWCNT content.[38] However, the CNT content of both of the samples in Fig. 5A are the same, and this is explained by sulfuric acid treatment. Sulfuric acid did not just change the surface composition of the samples but by degradation of some of the PDMS. Our FTIR data showed similar results as previous study that conducted by Gitlin et al. who demonstrated that sulfuric acid treatment on the PDMS surface was changed the atomic composition of the surface. Their results showed that SO$_4$ units incorporate as esterification reaction took place which was occurred after the cleavage reaction of Si-C bond.[36]

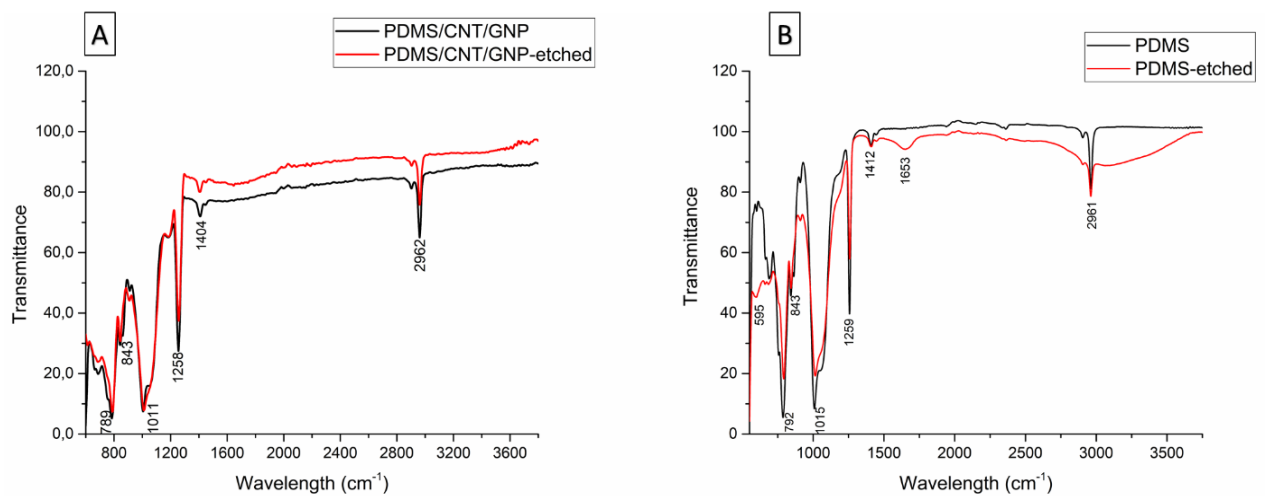

**Figure 5.** FTIR analysis result of the samples. A) FTIR results of untreated and H$_2$SO$_4$ treated PDMS/CNT/GNP nanocomposites. B) FTIR analysis results of untreated and H$_2$SO$_4$ treated PDMS samples

Table 1. Assignment of the ATR-FTIR bands

| Wavelength (cm$^{-1}$) | Interpretation |
|---|---|
| ~3000–3600 | OH stretching |
| ~2962 | Symmetrical CH$_3$ stretching |
| ~2908 | Asymmetrical CH$_3$ stretching |
| 1830–1560 | H$_2$O bending |
| ~1443 | CH$_2$ bending |
| ~1404 | CH$_3$ asymmetrical bending |
| ~1259 | Deformation vibration of -CH$_3$ in Si–CH$_3$ |
| 1176 | S=O symmetrical stretching. C=O stretching |

| ~1015 | Si–O–Si asymmmetrical deformation |
| 910 | C–O–S stretching |
| 864 | CH$_3$ symmetrical rocking |
| 843 | Si–C asymmetrical stretching |
| ~792 | CH$_3$ asymmetrical rocking |
| 756 | Si–C symmetrical stretching |
| 687 | Si–CH$_3$ symmetrical rocking |

Electrical conductivities of prepared samples were measured using the four-point probe method. Electrical conductivity result is shown in Fig. 6. We were expecting that the conductivity of PDMS61 would be higher than other samples without sulfuric acid treatment. Although previous experiments showed that increase in MWCNT:GNP ratio let to an increase in electrical conductivity.[14,25] However, our test results indicated that PDMS70 (278.3 S/m) and PDMS52 (630 S/m) samples were more conductive than PDMS61 (270.7 S/m). We measured the conductivity at various part of the nanocomposite samples, to show nanofillers were homogenously dispersed inside the PDMS matrix. The reason for lower electrical conductivity of PDMS61 might be that the nanofillers agglomerated due to poor mixing, or PDMS61 might not cure completely, in which the latter one is more possible. Because, sulfuric acid-treated samples (PDMS5s, PDMS10s, PDMS15s, and PDMS30s) were prepared from the same batch as PDMS61, and all of these samples were significantly more conductive than other samples. Moreover, PDMS5s and PDMS30s showed higher conductivity than PDMS10s and PDMS15s however, we were expected that the conductivities of the latter two samples would be higher than PDMS5s. This might be also due to the same reasons as previously mentioned. Sulfuric acid treatment affected the conductivity of the samples in a positive way. As mentioned before, sulfuric acid also affected the surface composition of the samples, and SO$_4$ units that incorporated into PDMS was changing the electrical conductivity of PDMS. As shown in Fig. 6, empty PDMS was not conductive but, PDMS-H2SO4 showed slight conductivity that was approximately $3.47 \times 10^{-1}$ S/m.

Previous studies were utilized solution mixing method to obtain highly conductive nanocomposites. Lu et al. demonstrated how high nanofiller concentration was affecting the electrical conductivity of nanocomposites. They obtained nanocomposite with conductivity value of 62,9 S/m when MWCNT concentration was 25 wt%, which is the highest value we found in the literature, but its viscosity is very high to be used in micro-fabrication. Moreover, they concluded that for the sensor applications, nanofiller concentration should be between 8-18 %.[8] Several studies were investigated the nanocomposites consisting of CNT and GNP as hybrid nanofillers, as shown in Table 2. These studies were prepared nanocomposites with nanofiller concentration of only 1 wt% and obtained better conductivity values when compared to other studies that only used one kind of nanofiller material.[14,25,35] When compared with the electrical conductivity results obtained from literature, our results showed significantly higher conductivity. Even in PDMS70, which has only MWCNTs as nanofiller material, we obtained better results than other studies shown in Table 2.

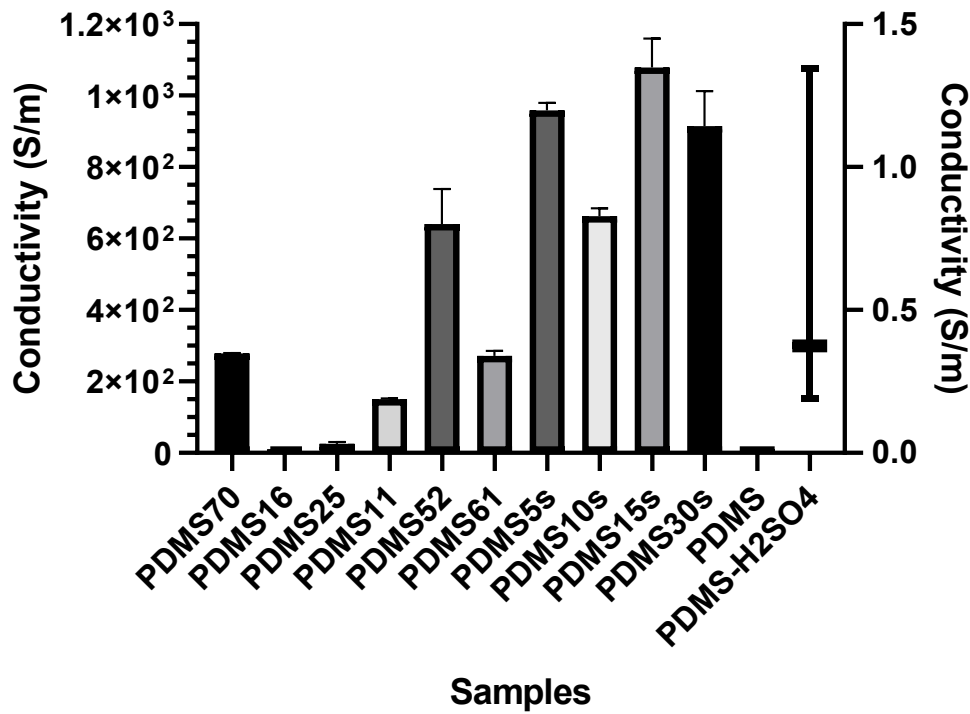

**Figure 6.** Electrical conductivity (S/m) values of different samples Conductivities (S/m) of nanocomposite samples shown in left y-axis, and conductivities (S/m) of PDMS and PDMS-H2SO4 shown in right y-axis)

**Table 2.** Comparison of electrical conductivity values obtained from the literature

| Polymer matrix | Nanofiller type | Polymer to nanofiller ratio | Production method | Electrical conductivity (S/m) | Ref. |
|---|---|---|---|---|---|
| PDMS | MWCNT GNP | 7 wt% | Solution mixing | 270,7 – 1074,8 | This work |
| PDMS | MWCNT | 8 wt% | Solution mixing | 9,9 | [5] |
| PDMS | MWCNT | 15 wt% | Solution mixing | 6,3 | [2] |
| PDMS | MWCNT | 2 vol% | Solution mixing | ~0,24 | [39] |
| PDMS | MWCNT-COOH | 4.5 wt% | Solution mixing | 23[a] | [40] |
| PDMS | MWCNT GNP | 4.5 wt% | Solution mixing | 3,4x$10^{-3a}$ | [41] |
| PDMS | MWCNT | 25 wt% | Solution mixing | 62,9[a] | [8] |
| PDMS | MWCNT | 12 wt% | Solution mixing | 34.9 | [42] |
| PDMS | MWCNT GNP | 1 wt% | Mixing | ~1 | [14] |

| PDMS | MWCNT GNP | 1 wt% | Solution mixing | 1,8[a] | 25 |
| PDMS | MWCNT GNP | 1 wt% | Solution mixing | 1,37 | 35 |

[a] Values extracted from graph.

## 4. Conclusion

In conclusion, we produced PDMS/MWCNT and PDMS/MWCNT/GNP nanocomposites with high electrical conductivity. However, the electrical conductivity of hybrid nanocomposites is modified by means of mixture ratio or/and type of nanofiller. We have observed that there is a synergetic effect at certain values in nanocomposites to get an increase conductivity thorough PDMS. We also observed that electrical conductivity of samples was greatly increased by sulfuric acid treatment onto the surface of nanocomposites, even empty PDMS showed slight conductivity. Additionally, FTIR data of the samples showed that sulfuric acid treatment was changed the surface composition of PDMS and nanocomposites. Electrical conductivity measurement results of nanocomposites were as high as 1074.8 S/m that is best value compared the results in the literature. There is an increase of conductivity by four order of magnitude compared to acid treated pristine PDMS sample. We have successfully obtained highly conductive PDMS based nanocomposite which could be utilized in microfluidics applications.

## Acknowledgements